\documentstyle[12pt,epsfig]{article}
\begin{document}
\vspace{1.cm}
\begin{center}
\    \par
\    \par
\    \par
\     \par

              {\bf{THE ANALYSIS OF $\pi^{-}$ MESONS PRODUCED
           IN NUCLEUS-NUCLEUS COLLISIONS AT A MOMENTUM OF 4.5
           GeV/c/NUCLEON IN LIGHT FRONT VARIABLES}}
\end{center}
\   \par
\   \par
\   \par
\   \par
\    \par
\par
    {\bf{ M.Anikina$^{a}$,    L.Chkhaidze$^{b}$,    T.Djobava$^{b}$,}}\par
    {\bf{ V.Garsevanishvili $^{c}$,     L.Kharkhelauri$^{b}$}}\par
\  \par
{\it{$^{a}$ Joint Institute for Nuclear Research, 141980 Dubna, Russia\par
$^{b}$ High Energy Physics Institute, Tbilisi State University,\par
   University Str. 9, 380086 Tbilisi, Republic of Georgia\par
$^{c}$ Mathematical Institute of the Georgian Academy of Sciences \par
   M.Alexidze Str. 1 , 380093 Tbilisi, Republic of Georgia }}
\   \par
\   \par
E-mail: djobava@sun20.hepi.edu.ge      \par
\pagebreak
\   \par
\   \par
\begin{center}

                     \bf{ ABSTRACT }
\end{center}
\ \par
\par
The light front analysis of $\pi^{-}$ mesons in He(Li,C), C-Ne, C-Cu
and O-Pb collisions is carried out. The phase space of secondary
pions is  divided into two parts in one of which the thermal
equilibrium assumption seems to be in a good agreement with the data.
Corresponding
temperatures  $T$ are extracted and their
dependence
on $(A_{P}*A_{T})^{1/2}$ is studied.
The results are compared with
the predictions of the Quark-Gluon String Model (QGSM).
The QGSM satisfactorily reproduces the experimental data
for light and intermediate-mass nuclei.
\par
\     \par
\    \par
\    \par
\     \par
PACS: 25.70.-z
\    \par
\    \par
{\it{Keywords}}: NUCLEAR REACTION He(Li,$\pi^{-}$,X), He(C,$\pi^{-}$,X),
C(Ne,$\pi^{-}$,X), C(Cu,$\pi^{-}$,\\X), O(Pb,$\pi^{-}$,X) at 4.5 GeV/c/nucleon;
measured pion disributions: deduced thermal equilibrium,
characterized by the temperature $T$. Analysis in light-front variables.
Comparison with the Quark Gluon String Model.
 Light-front variables; Phase space;
Thermal equilibrium.
\pagebreak
\begin{center}
\bf { 1.  INTRODUCTION }
\end{center}
\  \par
\par
In the present paper we continue the study of $\pi^{-}$ mesons produced
in the relativistic nucleus-nucleus collisions in terms of the light front
variables. The choice of the light front variables is due to the fact,
that these variables seem to be more sensitive to the dynamics of
interaction as compared to the well-known Feynman variables
$x_{F}$ and rapidity $Y$.
\par
In the previous publication [1] we performed the light front analysis of
$\pi^{-}$ mesons produced in Mg-Mg collisions at 4.3 GeV/c. On the basis
of this analysis we were able to separate the region of the
phase space, where the thermal equilibrium seems to be reached. It
is interesting in this connection to perform the same
analysis for other pairs of nuclei ( He(Li,C), C-Ne, C-Cu, O-Pb)
in order to study the dependence of the corresponding characteristics
on the atomic number of colliding pairs of nucleus . The data were obtained
on the
SKM-200-GIBS facility of the Dubna Joint Institute for Nuclear Research.
\  \par
\  \par
\begin{center}
\bf { 2.  EXPERIMENT }
\end{center}
\  \par
\par
  SKM-200-GIBS   consists  of  a   2m
streamer chamber, placed in a magnetic field of  $\sim$ 0.8  T  and   a
triggering system. The streamer chamber was exposed by beams  of
He , C , O ,  Ne  and  Mg  nuclei accelerated in the synchrophasotron  up
to a momentum of  4.5 GeV/c  per incident nucleon. The solid targets
in the form of thin discs with thickness  0.2$\div$0.4 g/cm$^{2}$
( for  Li the thickness was  1.59 g/cm$^{2}$ and for Mg 1.56 g/cm$^{2}$)
were mounted within the fiducial
volume
of the chamber. Neon gas filling of the chamber served also  as  a
nuclear target. The triggering system  allowed  the  selection  of
"inelastic"  and "central" collisions.
\par
   The central trigger was selecting events with no charged  and neutral
projectile
spectator fragments (with  $P/Z>3$ GeV/c ) within a cone of  half angle
$\Theta_{ch}$, $\Theta_{n}$ = 2$^{0}$ or  3$^{0}$.
The  trigger  mode  for each exposure is defined as  T ($\Theta_{ch}$
,$\Theta_{n}$ ).
  Thus   T(0,0)  corresponds to  all
inelastic  collisions.  For  inelastic  collisions He-Li and He-C
all charged secondaries were measured and central subsamples T(2,0)
were selected.
The details of the experimental setup and the logic of the triggering
systems are presented in [1,2].
\par
  Primary results of scanning and measurements were biased due to several
experimental effects and appropriate corrections were introduced. The biases
and correction procedures were discussed in detail in [2,3].
  Average measurement errors of the  momentum  and
production angle  determination for $\pi^{-}$   mesons were $<\Delta P/P
>$= 5$\%$, $\Delta$$\Theta$ =0.5$^{0}$ for He-Li, He-C, C-Ne, C-Cu, O-Pb.
\  \par
\  \par
\begin{center}
\bf { 3.  THE DATA ANALYSIS IN TERMS OF THE LIGHT FRONT VARIABLES }
\end{center}
\  \par
\par
The light front variables $\xi^{\pm}$
in the centre of mass frame for the inclusive reaction $a+b\to c+X$
are defined as follows [1]:
\begin{eqnarray}
\xi^{\pm}&=&\pm {E\pm p_z\over{\sqrt{s}}}=\pm {E+|p_z|\over{
\sqrt{s}}}  \label{eq1}
\end{eqnarray}
where $s$ is the usual Mandelstam variable,
$E=\sqrt{p^{2}_z+p^{2}_T+m^{2}}$ and $p_{z}$ are
the energy and the $z$ - component of the momentum of produced particle.
The $z$ -axis is taken to be the collision axis i.e. $p_{z}=p_{3}$.
The upper sign in Eq. (1) is used for the right hand side hemisphere and
the lower sign for the left hand side hemisphere.
In order to enlarge the scale in the region of small $\xi^{\pm}$,
 it is convenient also to
introduce the variables
\begin{eqnarray}
\zeta^{\pm}=\mp{\rm ln}|\xi^{\pm}| \label{eq2}
\end{eqnarray}

The invariant differential cross section in terms of these variables
looks as follows:
\begin{eqnarray}
E{d\sigma\over{d\vec p}}={|\xi^{\pm}|\over{\pi}}\ {d\sigma\over{
d\xi^{\pm}dp^{2}_T}} = {1\over{\pi}}\ {d\sigma\over{
d\zeta^{\pm}dp^{2}_T}}  \label{eq3}
\end{eqnarray}
\par
The light front variables have been introduced  by
Dirac [4] and they are widely used now in the treatment of
many theoretical problems
(see, e.g. original and review papers [5-10]).
\par
The analysis has been carried out for the $\pi^{-}$ mesons from
He(Li,C), C-Ne, C-Cu and O-Pb collisions in the nucleus-nucleus centre
of mass system.
\par
The number of events for all pairs of nuclei and corresponding trigger
modes are listed in Table 1.
Due to the small statistics and average multiplicities, the data of
He-Li and He-C collisions has been united and thus He(Li,C)
represents this sample of the data.
In Fig. 1   the $\xi^{\pm}$ -- distributions of $\pi^{-}$ mesons
from He(Li,C), C-Ne, C-Cu and O-Pb interactions
 are presented.  These distributions are similar for all analysed
pairs of nuclei.
The principal differences of $\xi^{\pm}$ distributions as compared to the
corresponding $x_{F}$ -- distributions  (Fig. 2) are the following:
(1) existence of some forbidden region around the point $\xi^{\pm}=0$;
(2) existence of maxima at some $\tilde{\xi^{\pm}}$ in the region of
 relatively small $|\xi^{\pm}|$.
\par
The experimental data
for invariant distributions $(1/\pi) \cdot dN/d\zeta^{\pm}$
are shown in Fig. 3. The curves are the result of
the polynomial
approximation of the experimental distributions.  The
maxima at $\tilde{\zeta}^{\pm}$ are also observed in the invariant
distributions $(1/\pi) \cdot dN/d\zeta^{\pm}$.
 However, the region
$|\xi^{\pm}|>|\tilde{\xi}^{\pm}|$ goes over to the region
$|\zeta^{\pm}|<|\tilde{\zeta}^{\pm}|$ and vice
versa (see Eqs. (1) and (2)).
The  value of maxima are observed
 at $\tilde{\zeta^{\pm}}=2.0\pm0.1$ for all pairs of nuclei.
The $\tilde{\zeta^{\pm}}$ is the function of the energy (see Eqs. (1), (2)) and
does not depend on the mass numbers of the projectile ($A_{P}$) and
target ($A_{T}$).
\par
In order to study the nature of these maxima we have divided the
phase space into two regions
$|\zeta^{\pm}|>|\tilde{\zeta}^{\pm}|$  ($\tilde{\zeta^{+}}$= 2.0)
and
$|\zeta^{\pm}|<|\tilde{\zeta}^{\pm}|$
and studied the
$p_{T}^{2}$ and the angular distributions of $\pi^{-}$ mesons in these regions
separetely.
The numbers of pions in these two regions are approximately
equal. For example in C-Cu interactions in the region
$|\zeta^{\pm}|>|\tilde{\zeta}^{\pm}|$ the number of pions is equal to
--1987 and in
$|\zeta^{\pm}|<|\tilde{\zeta}^{\pm}|$ --- 2212.
In Figs. 4, 5 the $p_{T}^{2}$ and the angular
 distributions of $\pi^{-}$ mesons
from  He(Li,C), C-Ne,
 C-Cu and O-Pb interactions in different regions of $\zeta^{+}$ ( $\zeta^{+} >
\tilde{\zeta^{+}}$
and $\zeta^{+} < \tilde{\zeta^{+}}$) in the forward hemisphere are presented.
\par
One can see from Figs. 4, 5, that the
$p_{T}^{2}$ and the angular distributions of $\pi^{-}$ mesons differ significantly in
$\zeta^{+} > \tilde{\zeta^{+}}$  and $\zeta^{+} < \tilde{\zeta^{+}}$
regions. The angular distribution of pions in the region $\zeta^{+} <
\tilde{\zeta^{+}}$ (Figs. 5.b and 5.c)
is sharply anisotropic in contrast to the almost flat distribution
in the region $\zeta^{+} > \tilde{\zeta^{+}}$ (Figs. 5.a and 5.c).
The flat behaviour of the angular distribution allows one to think that
one observes a partial thermal equilibrium  [11] in the region
 $|\zeta^{\pm}| > |\tilde{\zeta^{\pm}}|$ ($|\xi^{\pm}| < |\tilde{\xi^{\pm}}|$)
of phase space.
The slopes of
$p_{T}^{2}$ -- distributions differ greatly in different regions of
$\zeta^{\pm}$ (Fig. 4).
 Thus the values of $\tilde{\zeta^{\pm}}$
are the boundaries of the two regions with significantly
different characteristics of $\pi^{-}$ mesons. The validity of
this statement can be seen from the momentum distributions of
$\pi^{-}$ mesons in the Laboratory frame. Fig. 6 presents
the momentum disribution of pions from C-Cu collisions in the laboratory
frame. The shaded area corresponds to the  region
of $\zeta^{+} > \tilde{\zeta^{+}}$  and the non-shaded one to the region of
$\zeta^{+} < \tilde{\zeta^{+}}$. One can see from the Fig. 6,
that these two regions almost
 do not overlap in the momentum space unlike to the cms case
(overlap $\sim 45\%$).
The pions from the region $\zeta^{+} > \tilde{\zeta^{+}}$ have
small momentum, approximately up to 0.6 GeV/c as compared to the pions from
$\zeta^{+} < \tilde{\zeta^{+}}$
(the momentum of pions ranges from $\sim$ 0.6 GeV/c to 3 GeV/c). The
similar results have been also obtained for the other pairs of nuclei.
Fig. 7 presents the dependence of $<P>_{lab}$ on $\Theta_{lab}$ for
all analysed pairs of nuclei (He(Li,C) and C-Ne data are presented with
the same symbol because of the similarity of their dependences)
and Mg-Mg interactions
in the $\zeta^{+} > \tilde{\zeta^{+}}$
and $\zeta^{+} < \tilde{\zeta^{+}}$ regions. The shapes of these
dependences are different in two regions of $\zeta^{+}$. The curves are
the result of polynomial approximation. $<P>_{lab}$ decreases and
$<\Theta>_{lab}$ increases with the increasing of $A_{P}$,
$A_{T}$.
\par
To describe the spectra in the region $\zeta^{+} > \tilde{\zeta^{+}}$ the
Boltzmann
\begin{center}
$  f(E)\sim e^{-E/T} $
\end{center}
distribution have been used.
\par
 The distributions
$1/\pi \cdot  dN/d\zeta^{+}$, $dN/dp_{T}^{2}$, $dN/dcos\Theta$ look in
this region as follows~:
\begin{eqnarray}
{1\over{\pi}}\ {dN\over{d\zeta^+}}&\sim&\int_0^{p^2_{T,max}}Ef(E)dp^2_T
\label{eq4}\\
{dN\over{dp^2_T}}&\sim&\int_0^{p^{}_{z,max}}f(E)dp^{}_z \label{eq5}\\
{dN\over{d\cos\theta}}&\sim&\int_0^{p^{}_{max}}f(E)p^2dp \label{eq6}\\
E&=&\sqrt{\vec p\, ^2+m^2_{\pi}}\ ,\ \vec p\, ^2=p^2_z+p^2_T \label{eq7}
\end{eqnarray}
where:
\begin{center}
${p^{2}_{T,max}} =(\tilde{\xi^{+}}\sqrt{s})^{2} - m_{\pi}^{2}$
\end{center}
\begin{center}
${p_{z,max}} =[p_{T}^{2}+m^{2}-(\tilde{\xi^{+}}\sqrt{s})^{2}]/
(-2\tilde{\xi^{+}}\sqrt{s})$
\end{center}
\begin{center}
$p_{max}=(-\tilde{\xi^{+}}\sqrt{s}cos\Theta + \sqrt{(
\tilde{\xi^{+}}\sqrt{s})^{2}- m_{\pi}^{2}
sin^{2}\Theta})/sin^{2}\Theta$
\end{center}
\par
The experimental distributions in the region $\zeta^{+} >\tilde{\zeta^{+}}$
have been fitted by the expressions (4), (5), (6), respectively. The results
of the joint fit of the distributions
$1/\pi \cdot  dN/d\zeta^{+}$, $dN/dp_{T}^{2}$, $dN/dcos\Theta$
are given in Table 1 and Figs. 4, 5, 8.
They show a rather good agreement with experiment.
 In Table 1 the values of the parameter $T$ obtained by fitting the data
with Boltzmann distribution are presented.
In order to determine how the characteristics vary the analysis
have been carried out also for $\tilde{\zeta^{+}}$=1.9 and 2.1.
The results are similar, but the joint fit of the distributions is
better for $\tilde{\zeta^{+}}$=2.0 (presented on Figures).
\par
 The spectra of $\pi^{-}$ mesons in the region $\zeta^{+} >
\tilde{\zeta^{+}}$ are
satisfactorily described by the formulae which follow from
the thermal equilibrium. The same formulae when extrapolated to the region
$\zeta^{+} < \tilde{\zeta^{+}}$ (Fig. 8)
deviate significantly from the data. Therefore
in the region $\zeta^{+} < \tilde{\zeta^{+}}$ the $p_{T}^{2}$ -- distributions
has been fitted by the formula
\begin{eqnarray}
\frac {dN}{dp_{T}^{2}} \sim \alpha \cdot e^{-\beta_{1}P_{T}^{2}} +
(1-\alpha) \cdot e^{-\beta_{2}p_{T}^{2}} \label{eq8}
\end{eqnarray}
and the $\zeta^{+}$ -- distributions by the formula
\begin{eqnarray}
\frac{1}{\pi}\cdot\frac {dN}{d\zeta^{+}} \sim (1 - \xi^{+})^{n}=
(1 - e^{-\vert \zeta^{+}\vert})^{n} \label{eq9}
\end{eqnarray}
which is an analogue of the $(1-x_{F})^{n}$ dependence -- the result of
the well-known quark-parton model consideration (see, e.g. [12]), which for
$\pi^{-}$ mesons gives the value n=3.  The dependence
$(1 - e^{-\vert \zeta^{+}\vert})^{n}$ is in good agreement with experiment
in the region $\zeta^{+} < \tilde{\zeta^{+}}$ and deviates from it in the
region $\zeta^{+} > \tilde{\zeta^{+}}$ ( Fig. 8).
 The results of the fit are given in
Table 1 and Figs. 4 and 8.
\par
Thus in the $\zeta^{\pm}$ ($\xi^{\pm}$) distributions we have singled out
points $\tilde{\zeta^{\pm}}$ ($\tilde{\xi^{\pm}}$) which seperate in the phase
space two groups of particles with significantly different characteristics.
There are no such points in the $x_{F}$ and $Y$ - distributions.
\par
 In this paper the Quark Gluon String Model
(QGSM) [13] is used for a comparison with experimental data. The QGSM is based
on the Regge and string phenomenology of particle production in inelastic
binary hadron collisions [14].
The QGSM simplifies the nuclear effects (neglects the potential
interactions between hadrons, Pauli blocking, coalescence of nucleons and etc.)
and concentrates on hadron rescatterings. The QGSM includes only low lying vector
mesons and baryons with spin 3/2, mostly $\Delta$ (3/2, 3/2) resonances.
A detailed description and
comparison of the QGSM with experimental data in a wide energy range can be
found in Ref. [1,15,16].
\par
We have generated He(Li,C), C-Ne, C-Cu, O-Pb interactions using Monte-Carlo
generator
COLLI, based on the QGSM. The events had been traced through the detector
and trigger filter.
\par
In the generator COLLI there are two possibilities to generate events:
1) at not fixed impact parameter $\tilde{b}$ and 2) at fixed $b$.
The events have been generated for $\tilde{b}$.
   From the impact parameter distribution
 we obtained the mean value of $<b>$.
For the obtained value of $<b>$, we have generated
a total samples of $A_{P}-A_{T}$  events. The numbers of generated
events for all analysed pairs of nuclei are listed in Table 1.
The two regimes
are consistent and it seems, that in our experiment the following
values of b are most probable: b=1.55 fm for He(Li,C);
b=2.20 fm for C-Ne; b=2.75 fm for C-Cu; b=3.75  fm for O-Pb.
\par
The experimental results have been compared with the predictions of the QGSM
 for the above mentioned values of $b$ and satisfactory  agreement between
 the experimental data and the model have been found.
In Figs 1.b  and 3.b the $\xi^{\pm}$ and $\zeta^{\pm}$ -
distributions
of $\pi^{-}$ mesons from the QGSM calculations are presented
together with the experimental ones for C-Cu interactions.
One can see, that the QGSM reproduces
these distributions well. The similar results have been obtained
for all analysed pairs of nuclei.
The QGSM also reproduces the $ p_{T}^{2} $  and $ cos\Theta $
distributions (Figs. 4.c and 5.c). The QGSM data show the similar characteristics
in the different regions of $\zeta$ as experimental ones:
 sharply anisotropic angular distributions in the region $\zeta^{+} <
\tilde{\zeta^{+}}$  and the almost flat distribution
in the region $\zeta^{+} > \tilde{\zeta^{+}}$; the slopes of
$p_{T}^{2}$ -- distributions differ greatly in different regions of
$\zeta^{+}$; the momentum distributions of pions in the laboratory
frame in different regions of $\zeta^{+}$ have the similar different shape
of spectra as experimental ones (Fig 6). Momentum distributions of QGSM
data reproduces the corresponding experimental spactra in both regions of
$\zeta^{+}$.
The distributions obtained by the
 QGSM in the region $\zeta^{+} >\tilde{\zeta^{+}}$
have been fitted by the expressions (4), (5), (6). The results
of the fit are given in Table 1 and Figs. 4.c, 5.c.
In the region $\zeta^{+} < \tilde{\zeta^{+}}$ the $p_{T}^{2}$
and the $\zeta^{+}$ -- distributions
have been fitted by the formulae (8) and (9), respectively.
 The results of the fit are given in
Table 1 and Fig. 4.c.
One can see from the Table 1, that
the values of the $T$ extracted from the experimental and QGSM data
coincide within the errors. The QGSM does not reproduce satisfactorily
the O-Pb data. This is may be caused by the fact, that QGSM simplifies
the nuclear effects, which are more pronounced for heavy nuclei.
In Ref. [17] it has been indicated that the model can be improved by including
higher mass baryon resonances and taking into account a possible increase
of the pion absorption cross section, $\sigma_{\Delta\Delta \rightarrow NN}$
in dense baryon medium, in comparison with the cross section, obtained
from detailed balance relation.
\par
In Fig. 9 the dependence of the parameter $T$  from the Table 1 on
$(A_{P}*A_{T})^{1/2}$, obtained from the experimental and QGSM data,
is presented. The temperature for Mg-Mg interactions is extracted from [1].
One can see, that $T$ decreases
linearly with the increasing of $(A_{P}*A_{T})^{1/2}$ i.e with the increasing
number of participating nucleons. Similar behaviour is predicted by the QGSM.
\par
In our previous article [18] the temperatures of pions in He-Li, He-C,
C-Ne, C-Cu and O-Pb interactions were obtained be means of inclusive
kinetic energy and transverse momentum spectra in central rapidity
interval (0.5 -- 2.1 for light nuclei and 0.1 -- 1.8 for heavy ones),
which corresponds to the pionization region and with the c.m.s. angles
$90 ^{0} \pm 10^{0}$. The pion spectra for He-Li, He-C and C-Ne
have been fitted by one exponent and for C-Cu and O-Pb by
a sum of two exponents, or two temperatures $T_{1}$
 and $T_{2}$ (describing the low and high momentum part of the spectrum).
 The temperatures extracted by the light front
analysis for light pairs of nuclei are less  about (15--20) $\%$ as
compared to those obtained in Ref. [18]. For heavy pairs of nuclei the
temperatures are more close to the low temperature $T_{1}$. It seems
obvious, as the thermal equilibrium region corresponds to
lower momenta.
It should be
mentioned that the extraction procedures of $T$ in the light-front
variables and in Ref. [18] are quite different and it seems, that different
regions of phase space are seperated by these two methods.
\par
It is interesting to compare the temperatures extracted by means of light
front analysis with those obtained in the GSI experiments
(FOPI, KAON and TAPS-
Collaborations, see, e.g. [19,20]).  The $T$ in the GSI experiments
have been obtained in a same manner as in our Ref. [18].
The numerical values of the parameter
$T$ for pions in Au-Au collisions at 1 A GeV and our values for heaviest
colliding pair are close to each other.
It seems interesting in this connection
to perform the light front analysis of the GSI, AGS and SPS data.
\  \par
\  \par
\begin{center}
\bf { 4.  CONCLUSIONS }
\end{center}
\  \par
\par
The light front
analysis of $\pi^{-}$ -- mesons in the relativistic nucleus-nucleus
collisions is carried out. The results of this paper confirm the conclusions
of our previous publication [1], that the phase space of secondary
$\pi^{-}$ mesons is divided into two parts, in one of which the thermal
equilibrium seems to be reached. Corresponding temperatues are
obtained from the fitting of the data by the Boltzmann distribution.
The characteristics of the $\pi^{-}$ mesons (the momentum, angular,
$p_{T}^{2}$ -- distributions) in these two regions differ significantly.
The dependence of $T$ on $(A_{P}*A_{T})^{1/2}$  has been studied. The
temperature decreases with increase of $(A_{P}*A_{T})^{1/2}$. The
experimental results have been compared with the QGSM. The model seems to be
in a good agreement with the data excluding O-Pb collisions -- the
heaviest colliding pair.
\   \par
\   \par
\   \par
\   \par
\   \par
ACKNOWLEDGEMENTS
\   \par
\   \par
\par
The authors express their deep gratitude to J.-P.Alard,
 Sh.Esakia, D.Ferenc, G.Kuratashvili,
J.-F.Mathiot, Z.Menteshashvili, G.Paic
for interesting discussions.
\pagebreak
\   \par
\   \par
\   \par

\newpage
\begin{center}
\bf{FIGURE CAPTIONS}
\end{center}
\  \par
{\bf{Fig. 1}}
{The  $\xi^{\pm}$ -- distribution of  $\pi^{-}$ mesons~~
a) from $\ast$ -- He(Li,C),
$\triangle$ -- C-Ne,  $\circ$   --  O-Pb
interactions. b) From  C-Cu interactions:
 $\circ$   --  the experimental data,  $\triangle$ -- the QGSM data.\\
The curves are the result of polynomial approximation
of the experimental data.}
\par
\  \par
{\bf{Fig. 2}}
{The  $x_{F}$ -- distribution of  $\pi^{-}$ mesons
from $\ast$ -- He(Li,C),
$\triangle$ -- C-Cu,  $\circ$   --  O-Pb
interactions.}
\par
\   \par
{\bf{Fig. 3}}
{The  $ \zeta^{\pm} $ -- distribution of  $\pi^{-}$ mesons ~~
a) from  $\ast$ -- He(Li,C), $\triangle$ -- C-Ne,  $\circ$   --  O-Pb
interactions.
b) From  C-Cu interactions:
 $\circ$   --  the experimental data,  $\triangle$ -- the QGSM data.\\
The curves are the result of polynomial approximation
of the experimental data.}
\par
\  \par
{\bf{Fig. 4}}
{The $ p_{T}^{2} $ distribution of  $\pi^{-}$ mesons
 from $\ast$ -- He(Li,C), $\triangle$ -- C-Ne,  $\circ$   --  O-Pb
interactions ~~
a) for $\zeta^{+} > \tilde{\zeta^{+}}$.~~
b) For $\zeta^{+} < \tilde{\zeta^{+}}$.~~
c) From C-Cu interactions for $\zeta^{+} > \tilde{\zeta^{+}}$
$\circ$   -- experimental and
$\bigtriangleup$ -- the QGSM data; for $\zeta^{+} < \tilde{\zeta^{+}}$
\mbox{\put(3.,0.){\framebox(6.,6.)[cc]{}}}~~~~ -- experimental
and  $\ast$ -- the QGSM data.
\\
 The solid lines - fit of the
experimental data in the regions $\zeta^{+} > \tilde{\zeta^{+}}$
and
$\zeta^{+} < \tilde{\zeta^{+}}$
by the Eqs.(5) and (8), correspondingly.}
\par
\   \par
{\bf{Fig. 5}}
{The $ cos\Theta $ distribution of  $\pi^{-}$ mesons
from $\ast$ -- He(Li,C), $\triangle$ -- C-Ne,  $\circ$   --  O-Pb
interactions ~~
a) for $\zeta^{+} > \tilde{\zeta^{+}}$.~~
b) For $\zeta^{+} < \tilde{\zeta^{+}}$.~~
c) From C-Cu interactions
for $\zeta^{+} > \tilde{\zeta^{+}}$
$\circ$   -- experimental and
$\bigtriangleup$ -- the QGSM data;
for $\zeta^{+} < \tilde{\zeta^{+}}$,
\mbox{\put(3.,0.){\framebox(6.,6.)[cc]{}}}~~~~ -- experimental and
$\ast$ -- the QGSM data.
\\
 The solid lines - fit of the
experimental data in the region $\zeta^{+} > \tilde{\zeta^{+}}$
by the Eq.(6) and in the
$\zeta^{+} < \tilde{\zeta^{+}}$
by the polynom.}
\par
\   \par
{\bf{Fig. 6}}
{The momentum  distribution of  $\pi^{-}$ mesons
from C-Cu interactions in the laboratory system.
The shaded area corresponds to the region of
 $\zeta^{+} > \tilde{\zeta^{+}}$.}
\par
\   \par
{\bf{Fig. 7}}
{The dependence of $<P>_{lab}$ on $\Theta_{lab}$ in the regions
$\zeta^{+} > \tilde{\zeta^{+}}$ (bottom data) and
$\zeta^{+} < \tilde{\zeta^{+}}$ (top data)
for
\mbox{\put(3.,0.){\framebox(6.,6.)[cc]{}}}~~~~ --  He(Li,C) and C-Ne,
$\diamond$ -- C-Cu,
$\ast$ -- Mg-Mg,
$\circ$ --  O-Pb.
The curves -- result of polynomial approximation.}
\par
\   \par
{\bf{Fig. 8}}
{The  $ (1/\pi) \cdot dN/d\zeta^{+}  $ distribution of  $\pi^{-}$ mesons
from C-Cu interactions.
$\circ$ -- experimental data, the solid line -- fit of the experimental
data in the region
$\zeta^{+} > \tilde{\zeta^{+}}$
by the Eq.(4),
 the dashed line --
fit of the experimental data in the region
$\zeta^{+} < \tilde{\zeta^{+}}$  by the Eq.(9).}
\par
\   \par
{\bf{Fig. 9}}
 {The dependence of the  parameter $T$ on $(A_{P}*A_{T})^{1/2}$
for  He(Li,C), C-Ne, Mg-Mg, C-Cu and O-Pb. $\circ$ -- the experimental data,
$\bigtriangleup$ -- the QGSM data.
The dashed line is a
result of linear approximation of the experimental data.}
\   \par
\   \par
\   \par
\   \par
\begin{center}
\bf{TABLE CAPTIONS}
\end{center}
\   \par
\   \par
{\bf{Table 1.}}
{ Number of the events, trigger
and the
results of the joint fit of the distributions
$1/\pi \cdot  dN/d\zeta^{+}$, $dN/dp_{T}^{2}$, $dN/dcos\Theta$
of $\pi^{-}$ --
mesons by Eqs. (4), (5), (6) in the region $\zeta^{+} >\tilde{\zeta^{+}}$
and $1/\pi \cdot  dN/d\zeta^{+}$ distributions by Eq.(9)
in the region $\zeta^{+} <\tilde{\zeta^{+}}$.}
\newpage
{\bf{Table 1.}} Number of the events, trigger
and the
results of the joint fit of the distributions
$1/\pi \cdot  dN/d\zeta^{+}$, $dN/dp_{T}^{2}$, $dN/dcos\Theta$
of $\pi^{-}$ --
mesons by Eq. (4), (5), (6) in the region $\zeta^{+} >\tilde{\zeta^{+}}$
and $1/\pi \cdot  dN/d\zeta^{+}$ distributions by Eq.(9)
in the region $\zeta^{+} <\tilde{\zeta^{+}}$.\\
\  \par
\  \par
\begin{tabular}{|c|c|c|c|c|}    \hline
 &  &  &  &  \\
$A_{p} - A_{T} $ &  & Number of & T~ (MeV)& $n$\\
$T(\Theta_{ch},\Theta_{n}$) &  & events & $\zeta^{+} >\tilde{\zeta^{+}}$ &
$\zeta^{+} <\tilde{\zeta^{+}}$ \\
 &  &  &  &  \\
\hline
 $ He(Li,C) $ &exp.& 6147 &81 $\pm$ 2 & 3.6 $\pm$  0.2\\
\cline{2-5}
$T(2,0)$ &QGSM&15566 &84 $\pm$ 2 & 3.5 $\pm$  0.1\\
\hline
 $ C-Ne $ &exp.& 902 &79 $\pm$ 3 & 3.7 $\pm$  0.2\\
\cline{2-5}
$T(2,0)$ &QGSM&3950 &82 $\pm$ 2 &3.4 $\pm$  0.8\\
\hline
 $ C-Cu $ &exp.&1203 &72 $\pm$ 2 & 3.0 $\pm$  0.1\\
\cline{2-5}
$T(3,3)$ &QGSM&3463 &74 $\pm$ 2 & 3.2 $\pm$  0.8\\
\hline
 $ O-Pb $ &exp.&732 &55 $\pm$ 3 & 2.6 $\pm$  0.1\\
\hline
\end{tabular}
\newpage
\begin{figure}
\begin{center}
\epsfig{file=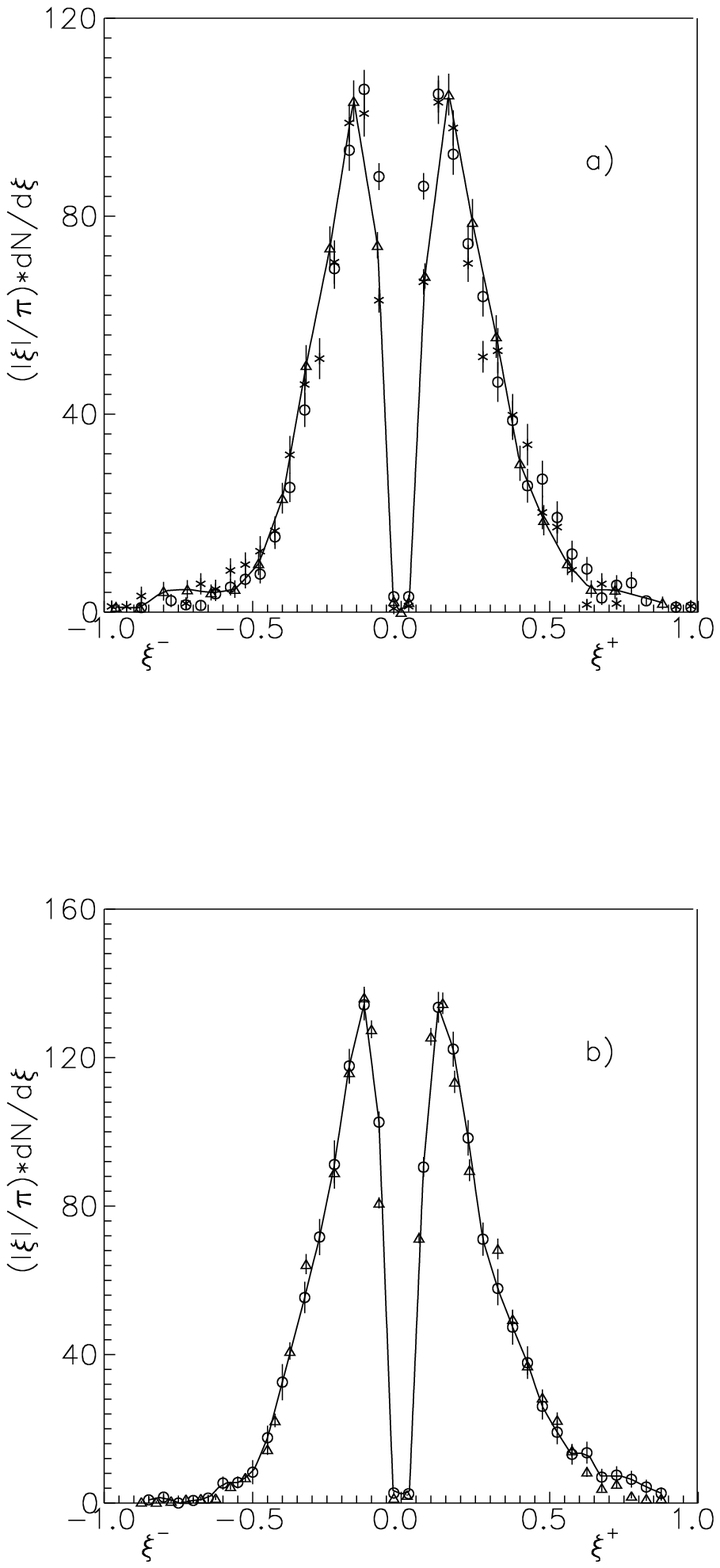,bbllx=0pt,bblly=0pt,bburx=594pt,bbury=842pt,
width=18cm,angle=0}
\end{center}
\vspace{-6.4cm}
\hspace{0.cm}
\begin{minipage}{16.0cm}
\caption
{The  $\xi^{\pm}$ -- distribution of  $\pi^{-}$ mesons~~
a) from $\ast$ -- He(Li,C),
$\triangle$ -- C-Ne,  $\circ$   --  O-Pb
interactions. b) From  C-Cu interactions:
 $\circ$   --  the experimental data,  $\triangle$ -- the QGSM data.
The curves are the result of polynomial approximation
of the experimental data.}
\end{minipage}
\end{figure}
\newpage
\begin{figure}
\begin{center}
\epsfig{file=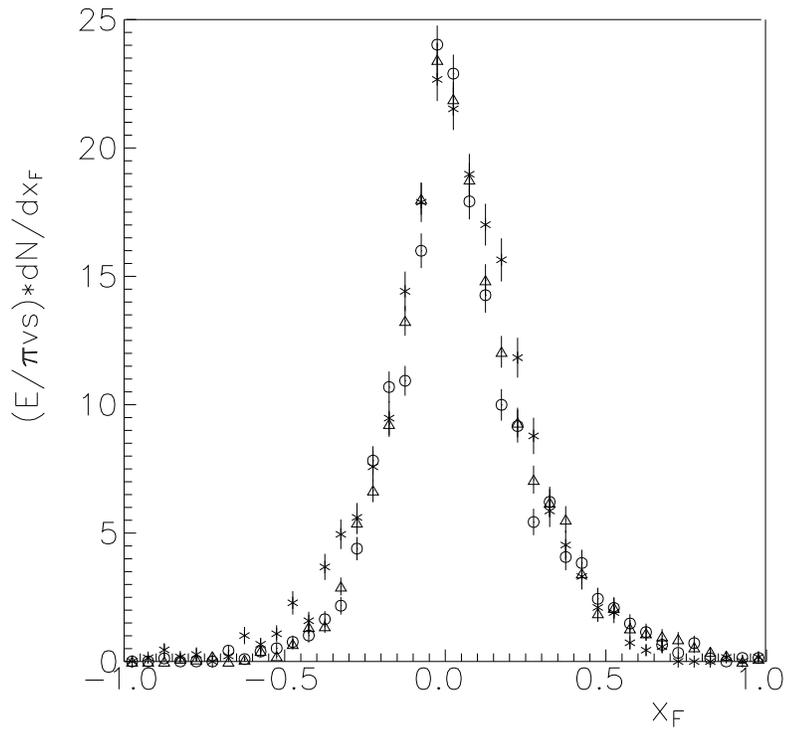,bbllx=0pt,bblly=0pt,bburx=594pt,bbury=842pt,
width=18.0cm,angle=0}
\end{center}
\vspace{-10.cm}
\hspace{3.cm}
\begin{minipage}{10.0cm}
\caption
{The  $x_{F}$ -- distribution of  $\pi^{-}$ mesons
from $\ast$ -- He(Li,C),
$\triangle$ -- C-Cu,  $\circ$   --  O-Pb
interactions.}
\end{minipage}
\end{figure}
\newpage
\begin{figure}
\begin{center}
\epsfig{file=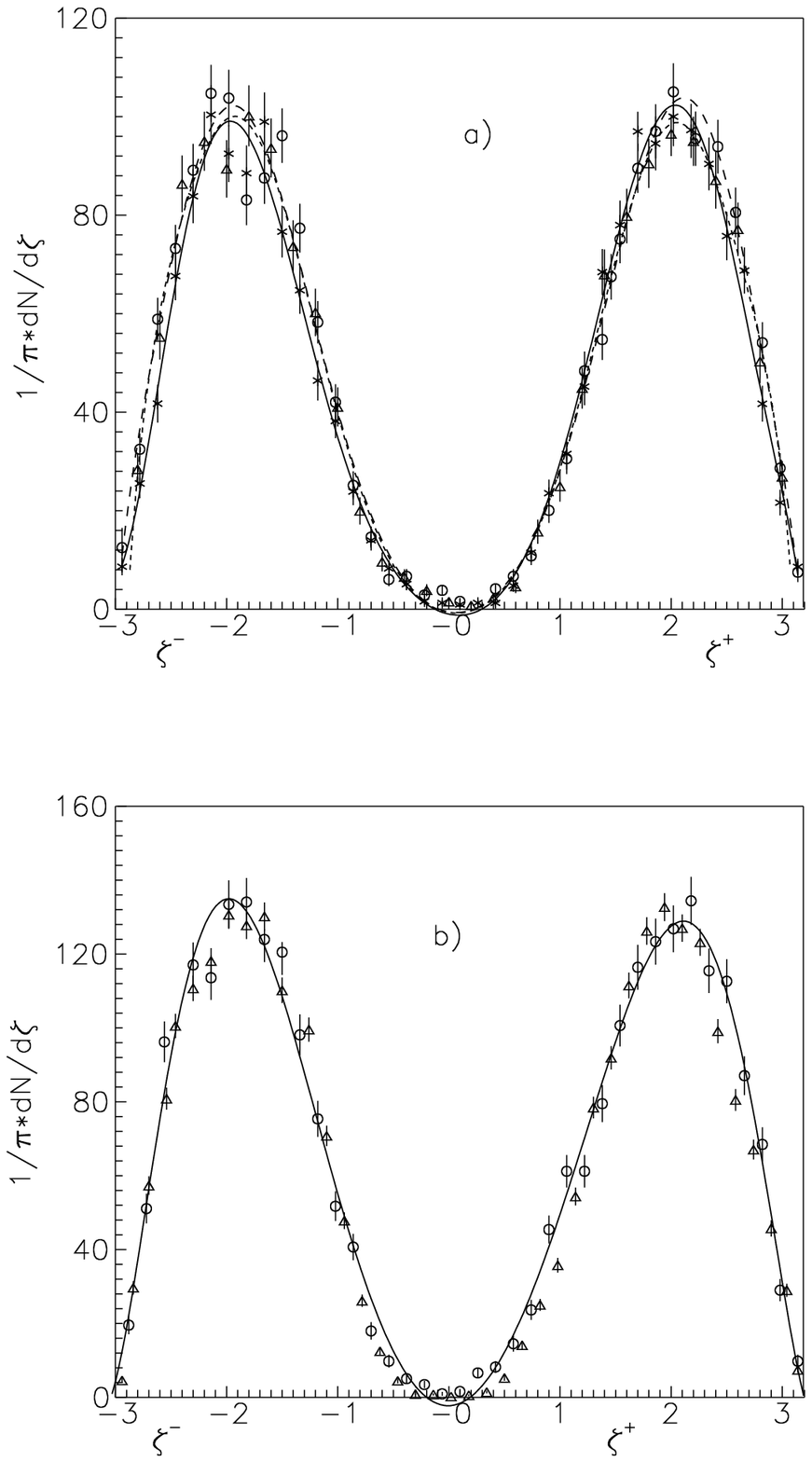,bbllx=0pt,bblly=0pt,bburx=594pt,bbury=842pt,
width=18cm,angle=0}
\end{center}
\vspace{-6.4cm}
\hspace{0.cm}
\begin{minipage}{16.0cm}
\caption
{The  $ \zeta^{\pm} $ -- distribution of  $\pi^{-}$ mesons ~~
a) from  $\ast$ -- He(Li,C), $\triangle$ -- C-Ne,  $\circ$   --  O-Pb
interactions.
b) From  C-Cu interactions:
 $\circ$   --  the experimental data,  $\triangle$ -- the QGSM data.
The curves are the result of polynomial approximation
of the experimental data.}
\end{minipage}
\end{figure}
\newpage
\begin{figure}
\begin{center}
\epsfig{file=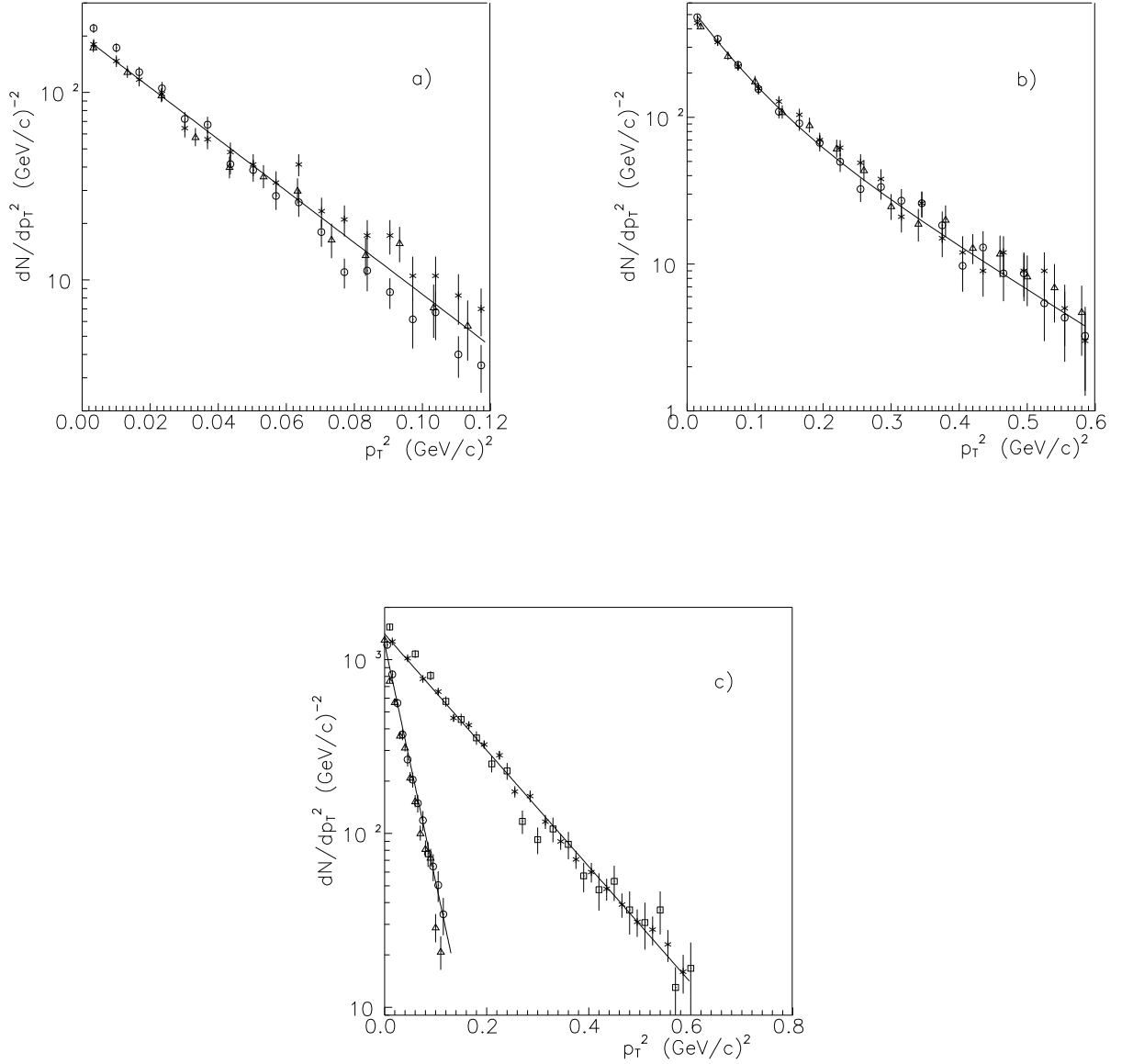,bbllx=0pt,bblly=0pt,bburx=594pt,bbury=842pt,
width=18cm,angle=0}
\end{center}
\vspace{-7.3cm}
\hspace{0.cm}
\begin{minipage}{16.0cm}
\caption
{The $ p_{T}^{2} $ distribution of  $\pi^{-}$ mesons
 from $\ast$ -- He(Li,C), $\triangle$ -- C-Ne,  $\circ$   --  O-Pb
interactions ~~
a) for $\zeta^{+} > \tilde{\zeta^{+}}$.~~
b) For $\zeta^{+} < \tilde{\zeta^{+}}$.~~
c) From C-Cu interactions for $\zeta^{+} > \tilde{\zeta^{+}}$
$\circ$   -- experimental and
$\bigtriangleup$ -- the QGSM data; for $\zeta^{+} < \tilde{\zeta^{+}}$
$\Box$ -- experimental
and  $\ast$ -- the QGSM data.
 The solid lines - fit of the
experimental data in the regions $\zeta^{+} > \tilde{\zeta^{+}}$
and
$\zeta^{+} < \tilde{\zeta^{+}}$
by the Eqs.(5) and (8), correspondingly.}
\end{minipage}
\end{figure}
\newpage
\begin{figure}
\begin{center}
\epsfig{file=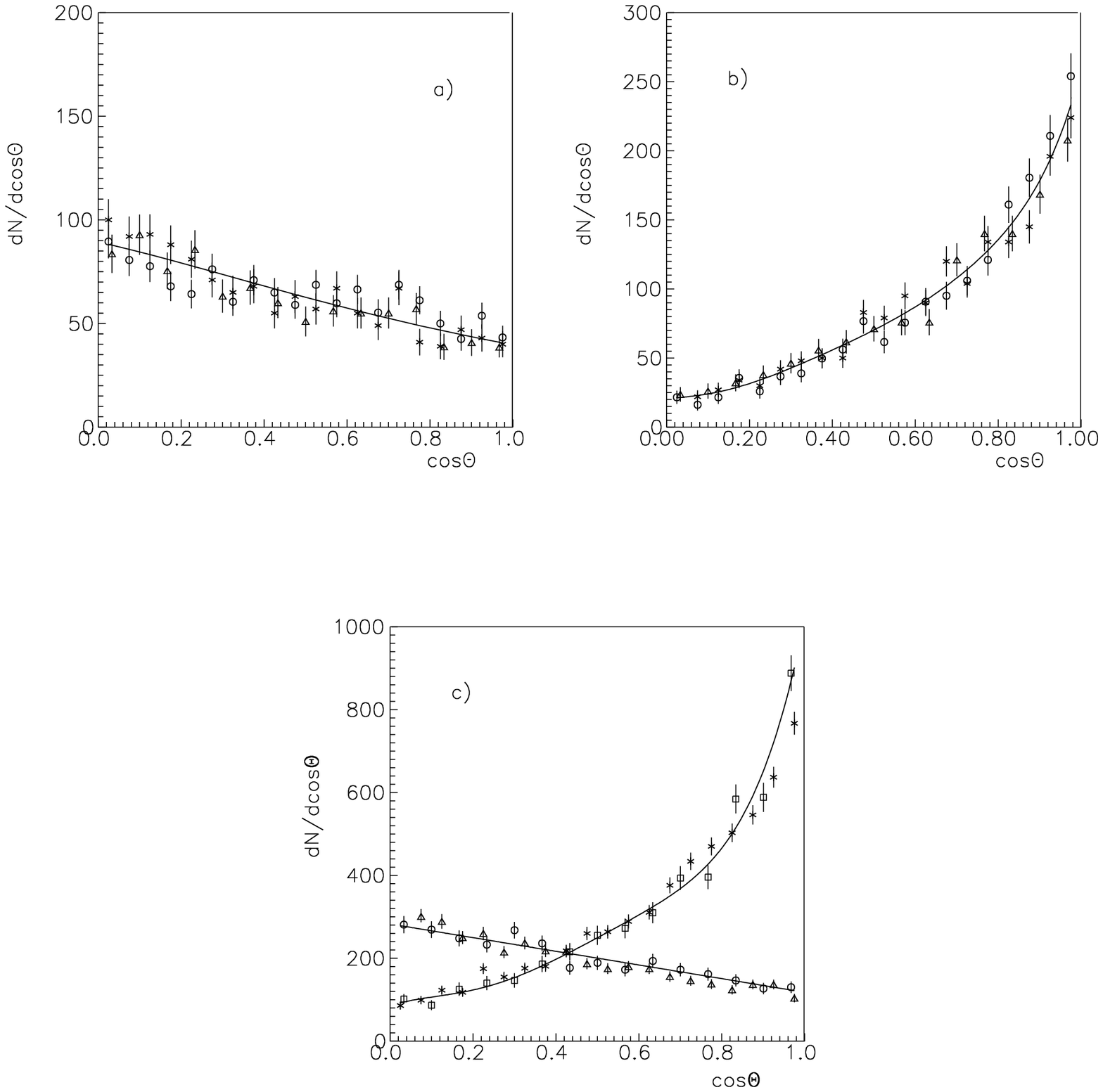,bbllx=0pt,bblly=0pt,bburx=594pt,bbury=842pt,
width=18cm,angle=0}
\end{center}
\vspace{-7.4cm}
\hspace{0.cm}
\begin{minipage}{16.0cm}
\caption
{The $ cos\Theta $ distribution of  $\pi^{-}$ mesons
from $\ast$ -- He(Li,C), $\triangle$ -- C-Ne,  $\circ$   --  O-Pb
interactions ~~
a) for $\zeta^{+} > \tilde{\zeta^{+}}$.~~
b) For $\zeta^{+} < \tilde{\zeta^{+}}$.~~
c) From C-Cu interactions
for $\zeta^{+} > \tilde{\zeta^{+}}$
$\circ$   -- experimental and
$\bigtriangleup$ -- the QGSM data;
for $\zeta^{+} < \tilde{\zeta^{+}}$
$\Box$ -- experimental and
$\ast$ -- the QGSM data.
 The solid lines - fit of the
experimental data in the region $\zeta^{+} > \tilde{\zeta^{+}}$
by the Eq.(6) and in the
$\zeta^{+} < \tilde{\zeta^{+}}$
by the polynom.}
\end{minipage}
\end{figure}
\newpage
\begin{figure}
\begin{center}
\epsfig{file=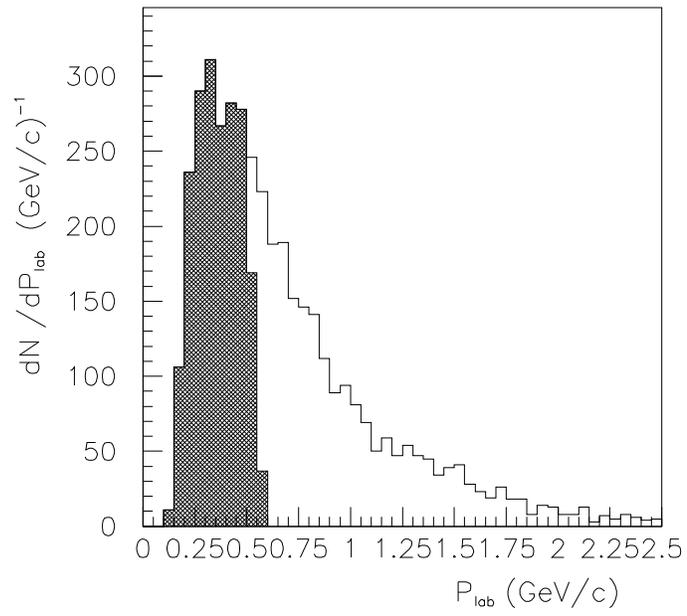,bbllx=0pt,bblly=0pt,bburx=594pt,bbury=842pt,
width=18cm,angle=0}
\end{center}
\vspace{-7.cm}
\hspace{0.cm}
\begin{minipage}{16.0cm}
\caption
{The momentum  distribution of  $\pi^{-}$ mesons
from C-Cu interactions in the laboratory system.
The shaded area corresponds to the region of
 $\zeta^{+} > \tilde{\zeta^{+}}$.}
\end{minipage}
\end{figure}
\newpage
\begin{figure}
\begin{center}
\epsfig{file=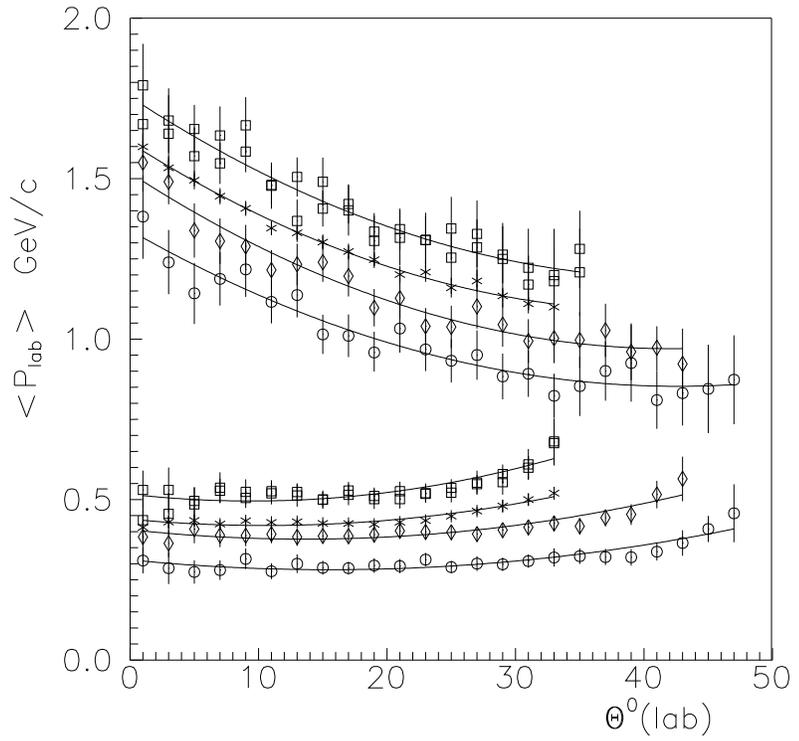,bbllx=0pt,bblly=0pt,bburx=594pt,bbury=842pt,
width=18cm,angle=0}
\end{center}
\vspace{-10.5cm}
\hspace{3.cm}
\begin{minipage}{12.0cm}
\caption
{The dependence of $<P>_{lab}$ on $\Theta_{lab}$ in the regions
$\zeta^{+} > \tilde{\zeta^{+}}$ (bottom data) and
$\zeta^{+} < \tilde{\zeta^{+}}$ (top data)
for
$\Box$ --  He(Li,C) and C-Ne,
$\diamond$ -- C-Cu,
$\ast$ -- Mg-Mg,
$\circ$ --  O-Pb.
The curves -- result of polynomial approximation.}
\end{minipage}
\end{figure}
\newpage
\begin{figure}
\begin{center}
\epsfig{file=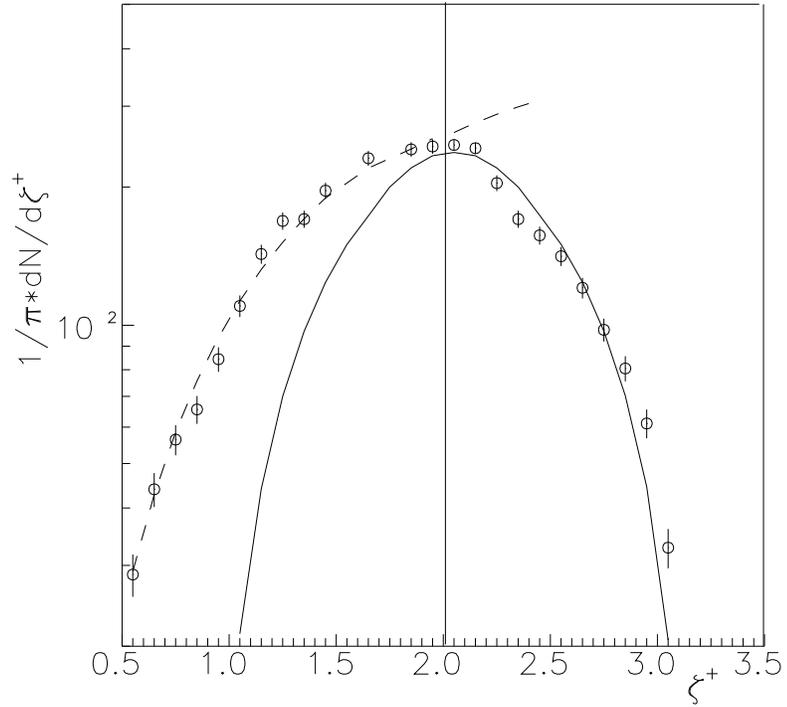,bbllx=0pt,bblly=0pt,bburx=594pt,bbury=842pt,
width=18cm,angle=0}
\end{center}
\vspace{-10.5cm}
\hspace{3.cm}
\begin{minipage}{12.0cm}
\caption
{The  $ (1/\pi) \cdot dN/d\zeta^{+}  $ distribution of  $\pi^{-}$ mesons
from C-Cu interactions.
$\circ$ -- experimental data, the solid line -- fit of the experimental
data in the region
$\zeta^{+} > \tilde{\zeta^{+}}$
by the Eq.(4),
 the dashed line --
fit of the experimental data in the region
$\zeta^{+} < \tilde{\zeta^{+}}$  by the Eq.(9).}
\end{minipage}
\end{figure}
\newpage
\begin{figure}
\begin{center}
\epsfig{file=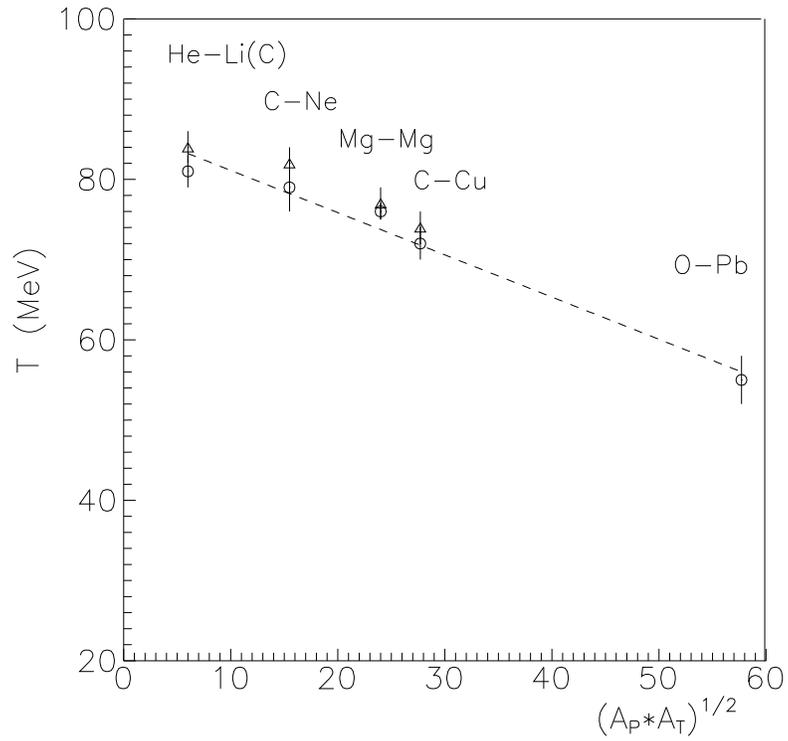,bbllx=0pt,bblly=0pt,bburx=594pt,bbury=842pt,
width=18cm,angle=0}
\end{center}
\vspace{-10.5cm}
\hspace{3.cm}
\begin{minipage}{12.0cm}
\caption
 {The dependence of the  parameter $T$ on $(A_{P}*A_{T})^{1/2}$
for  He(Li,C), C-Ne, Mg-Mg, C-Cu and O-Pb. $\circ$ -- the experimental data,
$\bigtriangleup$ -- the QGSM data.
The dashed line is a
result of linear approximation.}
\end{minipage}
\end{figure}
\end{document}